\documentclass[fleqn,twoside]{article}
\usepackage{espcrc2}

\usepackage{graphicx}
\usepackage[figuresright]{rotating}
\usepackage{amsmath}
\usepackage{amssymb}

\newcommand{\ie}{{\it i.e.}}

\newcommand{\eg}{{\it e.g.}}

\newcommand{\fig}{Fig.}

\newcommand{\Ref}{Ref.}
\newcommand{\Refs}{Refs.}

\newcommand{\stheta}{\sin^22\theta_{13}}
\newcommand{\deltacp}{\delta_\mathrm{CP}}
\newcommand{\ldm}{\Delta m_{31}^2}
\newcommand{\sdm}{\Delta m_{21}^2}

\newcommand{\figu}[1]{\fig~\ref{fig:#1}}

\newcommand{\bi}{\begin{itemize}}
\newcommand{\ei}{\end{itemize}}

\title{Long baseline neutrino oscillations: Theoretical aspects}

\author{Walter Winter\address{Institut f{\"u}r theoretische Physik und Astrophysik, Universit{\"a}t W{\"u}rzburg, D-97074 W{\"u}rzburg, Germany}%
        \thanks{Supported by the Emmy Noether program of Deutsche Forschungsgemeinschaft (DFG).}}
       
\begin{document}

\begin{abstract}
We discuss the measurement of the neutrino oscillation parameters at future long baseline experiments
in terms of the motivation of the experiments, the quantities of interest from the theoretical point of view, the phenomenology of these experiments, and the experiment choice. We illustrate the oscillation physics potential of a neutrino factory, as a representative
for the most challenging technologies. Finally, we point out that a future neutrino oscillation facility might also
be affected by the unexpected.

\end{abstract}

\maketitle

{\bf Theoretical motivation and quantities of interest.}
Future long baseline neutrino oscillation facilities are primarily designed to measure the unknown mixing angle $\theta_{13}$, for which only an upper bound exists, the neutrino mass ordering (mass hierarchy), and the leptonic CP phase $\deltacp$. In addition, there typically is good sensitivity to the atmospheric neutrino oscillation parameters $\ldm$ and $\theta_{23}$. But what motivates the measurement of these parameters?

The neutrino mixing parameters and mass squared differences are often assumed to originate from a theory beyond the Standard Model (BSM), see \Ref~\cite{Bandyopadhyay:2007kx} for an overview. Such a fundamental theory typically has two ingredients: It has to explain the smallness of neutrino mass compared to the charged leptons and quarks, and it has to describe the relative magnitude and ordering of the masses (flavor structure). The smallness of neutrino mass is typically assumed to come from integrating out a heavy BSM particle (type I, II, and III seesaw mechanisms), or from radiative corrections induced by such particles in loop diagrams. The flavor structure is often described by patterns for the Yukawa couplings (textures), discrete flavor symmetries, or anarchy arguments. A Grand Unified Theory (GUT) may, in addition, establish a connection between the quark and lepton sectors. 
Models from the discussed theories typically predict the neutrino oscillation observables, such as $\theta_{13}$ and the mass ordering  $\mathrm{sgn}(\ldm)$ (mass hierarchy), see, \eg, \Ref~\cite{Albright:2006cw}. Although the mass models can in most cases be not very precise in their predictions, since, for example, radiative corrections may change the results (see, \eg, \Ref~\cite{Antusch:2003kp}), it is obvious that improving the knowledge on these observables leads to a better understanding of the origin of neutrino mass. Therefore, the magnitude of $\theta_{13}$ and the mass hierarchy are good performance indicators for theoretical mass models. Other known performance indicators for theory are the sensitivity to CP violation as an indicator for leptogenesis (see, \eg, \Ref~\cite{Pascoli:2006ci} for a direct connection in certain cases), and the sensitivity to deviations from maximal mixing~\cite{Antusch:2004yx}.

Apart from these observables, there exist other discriminators among theoretical models. For example, in \Ref~\cite{Niehage:2008sg}, two different configurations with maximal mixings from both $U_\ell$ and $U_\nu$ have been considered ($\theta_{23}^\ell=\theta_{12}^\nu=\pi/4$ and the other angles zero), which lead to the following sum rules:
\begin{equation}
\theta_{12} \simeq \frac{\pi}{4} \pm \theta_{13} \, \cos \deltacp \, , \quad \theta_{23} = \frac{\pi}{4} \mp \frac{\theta_{13}^2}{2} \, ,
\end{equation}
where the upper signs stand for corrections from $\theta_{12}^\ell$, and the lower signs for corrections from $\theta_{13}^\ell$. From the first equation, we can read off that $\theta_{13}$ has to be large in both cases, and $\deltacp$ has to be close to $\pi$ or $0$, respectively, in order to obtain a small enough $\theta_{12}$ in the currently allowed range. From the second equation, we can moreover read off that the $\theta_{23}$ octant can discriminate the two possibilities, in principle. 
 Another example are relationships such as 
\begin{equation}
\theta_{12} + \theta_C \, \underbrace{\left( \frac{1}{\sqrt{2}} + \frac{\theta_C}{4} \right)}_{k} \simeq \frac{\pi}{4} 
\end{equation}
from \Ref~\cite{Plentinger:2008up}, which are representative for quark-lepton complementarity (QLC) as an indication for quark-lepton unification. Note that compared to the original QLC, as introduced in \Refs~\cite{Raidal:2004iw,Minakata:2004xt} which is incompatible with tribimaximal mixings~\cite{Frampton:2008ep}, $k\neq 1$ here, as it is typical for ``modern'' QLC approaches. Therefore, $k$ may serve as a performance indicator for quark-lepton unification. Finally, other performance indicators for theory could be $|\sin^2 \theta_{12} - 1/3|$ as indicator for tribimaximal mixings, and any deviation from the standard oscillation framework. 
In summary, there are a number of indicators which can discriminate different theoretical approaches. For long baseline experiments, the most important ones might be the magnitude of $\theta_{13}$, the mass hierarchy, leptonic CP violation, the value of $\deltacp$, deviations from maximal atmospheric mixing, and the $\theta_{23}$ octant.

{\bf Long baseline phenomenology.}
Long baseline experiments at the GeV scale produce neutrinos by the decays of pions and kaons (superbeams), muons (neutrino factories), or unstable nuclei (beta beams). The first oscillation channel of interest is the $\nu_\mu \rightarrow \nu_\mu$ disappearance channel\footnote{In this discussion, $\nu_\alpha$ can refer to antineutrinos as well.} to measure $\ldm$ and $\theta_{23}$. For $\theta_{13}$, the mass hierarchy, and $\deltacp$, one may use the ``golden''~\cite{DeRujula:1998hd,Cervera:2000kp} $\nu_e \rightarrow \nu_\mu$ (neutrino factory, beta beam) or $\nu_\mu \rightarrow \nu_e$ (superbeam) channel, the ``silver''~\cite{Donini:2002rm,Autiero:2003fu} $\nu_e \rightarrow \nu_\tau$ channel (neutrino factory), or the ``platinum''~\cite{Bandyopadhyay:2007kx} $\nu_\mu \rightarrow \nu_e$ channel (neutrino factory). The latter two possibilities are very challenging, since the silver channel typically operates with relatively low statistics (compared to the golden channel), and the platinum channel requires charge identification of the electrons. Other interesting channels are the $\nu_\mu \rightarrow \nu_\tau$ appearance channel to verify the oscillation into $\nu_\tau$'s (OPERA~\cite{Duchesneau:2002yq}, neutrino factory? -- see, \eg, \Ref~\cite{FernandezMartinez:2007ms}), and the neutral current measurement to study new physics~\cite{Barger:2004db,Adamson:2008jh}.

Let us now focus on the $\nu_e \leftrightarrow \nu_\mu$ oscillation channels to measure $\theta_{13}$, mass hierarchy, and $\deltacp$.  From analytical approximations of the oscillation probabilities, such as given 
in \Refs~\cite{Cervera:2000kp,Freund:2001pn,Akhmedov:2004ny} expanded to second order in $\theta_{13}$ and $\alpha \equiv \sdm/\ldm$, one can read off that these quantities of interest can, in principle, be extracted from the oscillation probabilities if the other oscillation parameters are known and if there is enough energy information to disentangle the different contributions to the oscillation probabilities. However, many experiments either have to poor energy resolution to disentangle this information, or have too narrow beam spectra to have energy information at all. Therefore, it is a good starting point to consider a total rate measurement first. In this case, one cannot extract $\theta_{13}$ and $\deltacp$ simultaneously, and one only obtains a curve in the $\theta_{13}$-$\deltacp$-plane reflecting the degenerate solutions. Therefore, it is a common strategy to use the CP-conjugated channels, such as $\nu_e \rightarrow \nu_\mu$ and  $\bar\nu_e \rightarrow \bar\nu_\mu$, which breaks the continuous degeneracy between $\theta_{13}$ and $\deltacp$. However, the Earth's matter potential is not CP-symmetric (because it does not contain antimatter), which means that matter effects obscure the extraction of fundamental CP violation. An oscillation channel which does not have this problem is the T-conjugated platinum channel, such as in a neutrino factory or by combining a superbeam with a beta beam.
However, note that for the mass hierarchy determination, the CP asymmetry of the matter profile can be used as a virtue. 
 
Even if both neutrinos and antineutrinos are used, there is several discrete
degeneracies remain in the parameter space: The
$(\delta, \theta_{13})$~\cite{Burguet-Castell:2001ez},
$\mathrm{sgn}(\Delta m_{31}^2)$~\cite{Minakata:2001qm}, and
$(\theta_{23},\pi/2-\theta_{23})$~\cite{Fogli:1996pv} degeneracies,
\ie, and overall ``eight-fold'' degeneracy~\cite{Barger:2001yr}. For the resolution
of correlations and degeneracies, a number of approaches have been discussed (see, \eg, \Ref~\cite{Bandyopadhyay:2007kx}):
{\bf Matter effects} to break the $\mathrm{sgn}(\Delta m_{31}^2)$-degeneracy and to measure the
mass hierarchy, such as by using a wide-band beam from Fermilab to an underground laboratory, in T2KK,
or a very long neutrino factory baseline.
{\bf  Different beam energies} or a better energy resolution at the detector, such as by using a wide-band beam
with a good enough detector, a beta beam with different isotopes, or a monochromatic beam.
{\bf 
 A second baseline,} such as a baseline from Japan to Korea for T2KK, a ``magic'' baseline~\cite{Huber:2003ak} at a neutrino factory or beta beam, or a second detector in the NuMI beam.
{\bf Better statistics,} such as by using a megaton-size detector, a neutrino factory, or a beta beam.
{\bf Additional channels,} such as by combining a superbeam and beta beam, using the golden and silver channels at a neutrino factory, or adding the information from reactor experiments.
{\bf Using other experiment classes,} such as atmospheric neutrino experiments.

An interesting discussion in this context is, for instance, the comparison between narrow-band
(off-axis) and wide-band (on-axis) superbeams. In \Ref~\cite{Barger:2007jq}, the optimization
for a NuMI-like beam pointed towards a 100~kt liquid argon detector has been performed in terms of 
off-axis angle and baseline. It has been demonstrated that, if the detector has good enough
energy resolution and background rejection, it is, in principle, better to be on-axis and
use the wide beam spectrum. It is currently a matter of discussion if a water Cherenkov detector
has the required quality in order to be used as such an on-axis detector. As another example,
the option to combine different baselines (and beam energies) has already been pointed out in the
context of the first-generation superbeams~\cite{Barger:2002xketc}:
In order to achieve complementarity between T2K and NO$\nu$A, a NO$\nu$A baseline significantly longer
than the MINOS baseline was needed.

{\bf Experiment choice and optimization.}
The choice of a future long baseline experiment will probably dependent on the outcome
of currently operated or constructed experiments, and other boundary conditions, such as
the LHC results. For the timescale, one may expect to know from reactor and long baseline
experiments that $\stheta \gtrsim 0.04$ by about 2011, and that $\stheta \gtrsim 0.01$ by 2015~\cite{Huber:2006vr}.
Obviously, choosing any of the two dates or $\stheta$ values as a decision point, one obtains
two qualitatively different cases relevant for the future experiment strategy:
$\theta_{13}$ discovered (``$\theta_{13}$ large'') and $\theta_{13}$ not discovered
(``$\theta_{13}$ small'').\footnote{Although the current best-fit for $\stheta$ is not
at zero anymore~\cite{Fogli:2008jx}, it does not affect this discussion since this deviation is not yet significant.}
These two cases are qualitatively different since for small $\theta_{13}$, one would
optimize for discovery reaches for as small as possible $\theta_{13}$, whereas for large
$\theta_{13}$, one would optimize for discovery reaches for as many as possible $\deltacp$.
We will illustrate these differences below.

Let us first of all discuss the large $\theta_{13}$ case, \ie, $\theta_{13}$ discovered.
For example, assume that Double Chooz establishes $\theta_{13} > 0 $ at  the 90\% confidence level.
In this case, there will be an allowed range for $\theta_{13}$, such as $0.06 \lesssim \stheta \lesssim 0.10$
for the best-fit $\stheta=0.08$ after three years of operators (1.5 years of these with a near detector)~\cite{Huber:2006vr}. For any future experiment, it is then easy to come up with a minimum wish list, which may look as follows~\cite{Winter:2008cn}:
\begin{itemize}
\item
$5 \sigma$ independent confirmation of $\theta_{13}>0$ for any (true) $\deltacp$,
\item
$3 \sigma$ mass hierarchy determination for any (true) $\deltacp$,
\item
$3 \sigma$ CP violation discovery for 80\% of all (true) $\deltacp$
\end{itemize}
for {\em any} possible $\theta_{13}$ in the remaining Double Chooz allowed range.
Any experiment which can fulfill these requirements is good enough
for the future neutrino program. However, one can also reverse the
problem: What is the {\em minimal} effort to have these sensitivities?
This minimal effort criterion has been studied in \Ref~\cite{Winter:2008cn}
for a beta beam. In this case, ``minimal'' refers to using only
one baseline at an as low as possible luminosity (affecting the ion source
and acceleration effort) and as low as possible boost factor (affecting the
acceleration effort). However, in a green-field scenario, the baseline can be
chosen arbitrarily without significantly affecting the effort. There are
two important outcomes of such a study: Within any realistic
scenario, the baseline has to be significantly longer than 500~km for the
mass hierarchy sensitivity, whereas the CP violation sensitivity typically
limits the minimal possible boost factor. For example, a boost factor as
low as 140 could be sufficient depending on the luminosity, baseline,
and $\theta_{13}$ best-fit, which means that a boost factor as high as~350
may not be necessary~\cite{BurguetCastell:2005pa}. The CERN-SPS may be
sufficient for this purpose if $^8$B and $^8$Li can be used as isotopes at
sufficiently high luminosities (about a factor of two higher than typically
anticipated).

For small $\theta_{13}$, \ie, $\theta_{13}$ not discovered, the situation
will be very different: One cannot clearly define criteria
such as the ones above. Since there has been no $\theta_{13}$ discovery,
it is unclear how small $\theta_{13}$ are actually good enough, and even
if one has decided, one has to make a choice for the fraction of
the (unknown) $\deltacp$ one wants to have this measurement for. As illustrated in
\Ref~\cite{Huber:2005jk} for beta beams, within realistic boundary conditions,
the minimal effort is essentially a matter of cost. 
The same applies to other types of long baseline experiments
which are not systematically limited. Therefore, it is quite likely that
the decision for a future long baseline facility will depend on other
impacts, such as the connection to the high energy frontier. For example,
the neutrino factory front-end may be shared with a muon collider test
facility~\cite{GeerVLC}. 

Finally, note that this discussion on experiment choice and optimization
has been purely conceptual. Especially for the next generation experiments,
the choice will depend on regional boundary conditions. These regional
perspectives are extensively discussed within this conference by a number
of other speakers.

{\bf Example: Neutrino factory.}
This section is devoted to the physics potential of 
a neutrino factory~\cite{Geer:1997iz} as an example for a future 
long baseline experiment. There was an International
Design Study~\cite{ids} (IDS-NF) initiated in 2007, which is an initiative
to present a design report, schedule, cost estimate and risk
assessment for a neutrino factory until about 2012. The IDS-NF
has defined a baseline setup based on the International Scoping Study
of a future neutrino factory and superbeam facility~\cite{Bandyopadhyay:2007kx,Berg:2008xx,Abe:2007bi}.
In short, this baseline setup uses two 50~kt magnetized iron detectors, one at 3000-5000~km 
and one at 7000-8000~km, and $10^{21}$
useful muon decays per year (in all polarities and baselines together),
operated at a muon energy of 25~GeV. The two-baseline optimization
for the main performance indicators has been revisited in \Ref~\cite{Kopp:2008ds},
where this combination has proven to be even unaffected by the presence of
 non-standard interactions (whereas the absolute sensitivities are deteriorated).
\begin{figure}[t]
\begin{center}
\includegraphics[width=0.8\columnwidth]{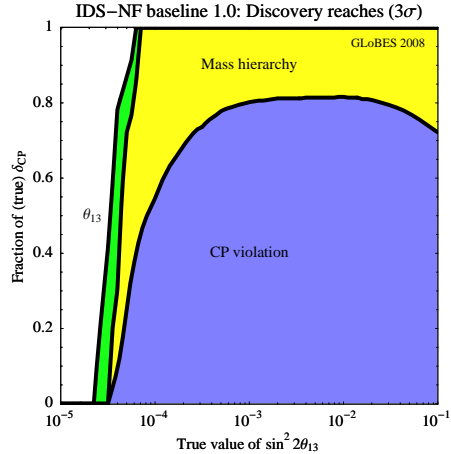}
\end{center}
\vspace*{-1cm}
\caption{\label{fig:allres}Discovery reaches for $\theta_{13}$, the (normal) mass hierarchy, and CP violation ($3\sigma$) for the IDS-NF baseline setup 1.0~\cite{ids} as a function of true $\stheta$ and fraction of (true) $\deltacp$. Discovery on right hand sides of the curves.}
\end{figure}
The discovery potential  for $\theta_{13}$, the (normal) mass hierarchy, and CP violation
for this setup is shown in \figu{allres}. It is quite impressive for the primary performance
indicators, which may be the main motivation to build this experiment. But what else
can we expect from a neutrino factory?

First of all, we expect precision for the quantities of interest. For example, it has been
shown in \Ref~\cite{Gandhi:2006gu} (Fig.~6) about a 10\% full width error on $\mathrm{log}_{10} \left( \stheta \right)$
can be obtained for the true $\stheta = 0.001$  ($3 \sigma$ CL). Note that this $\stheta$ will be about one order of magnitude smaller than the bound from the next generation of reactor experiments and superbeams.
Furthermore, as pointed out in \Ref~\cite{Huber:2004gg} (depending on the true $\deltacp$, see Fig.~7) about a 20 to 60 degree full width error can be achieved for $\deltacp$  for the true $\stheta = 0.001$  ($3 \sigma$ CL). But what does that mean? And what precision does one actually need from the theoretical point of view? For example, in certain QLC models, one can obtain sum rules such as~\cite{Niehage:2008sg}
\begin{equation}
\sin \deltacp \simeq \sqrt{2} \, \theta_C \, \sin 2 \Phi \, ,
\end{equation}
where $\Phi$ is a model parameter. Such sum rules or even systematic scans of the parameter
space~\cite{Winter:2007yi} motivate a Cabibbo-angle precision for $\deltacp$, \ie, about 13 degrees.
The above quoted full width error corresponds to such a precision.

Apart from the standard oscillation parameters, there is a number of other ``guaranteed'' oscillation
measurements at a neutrino factory. Depending on the magnitude of $\stheta$, one will obtain a matter density
measurement at the level of about 0.2\% ($\stheta=0.1$) to 2\% ($\stheta=0.001$) along the very long baseline at the $1\sigma$ CL~\cite{Gandhi:2006gu,Winter:2005we,Minakata:2006am},
which will be very complementary to seismic wave approaches. In addition, using the solar term in the appearance probability, one can verify the MSW effect in Earth matter at $5\sigma$ for $L \gtrsim 6000$~km even for $\stheta \equiv 0$~\cite{Winter:2004mt}. This  is very different from the mass hierarchy sensitivity, which is only rudimentarily present via the disappearance channel for $\stheta \equiv 0$~\cite{deGouvea:2005mi}. As a last example, for large enough deviations from maximal atmospheric mixing, one can resolve the octant degeneracy with the combination of the two baselines for any $\stheta$~\cite{Gandhi:2006gu}, or one can use the silver channel at the shorter baseline~\cite{Meloni:2008bd}.

{\bf The unexpected?}
During the planning of a future experiment or during operation, one should not forget that the
potentially unexpected could happen. 
For example, during operation, it may turn out that other new physics contributes to the measurements,
such as non-standard interactions (NSI) described by effective four fermion vertices. Such NSI may
be confused with the standard oscillation parameters~\cite{Huber:2002bi}, but also offer the chance
to discover new physics~\cite{Kopp:2007mi}.\footnote{The bounds expected from the IDS-NF setup can be found
in \Ref~\cite{Kopp:2008ds}.} As an interesting option, the NSI may even cause CP violation in the neutrino production/detection~\cite{FernandezMartinez:2007ms,Gonzalez-Garcia:2001mp,Altarelli:2008yr}  or propagation~\cite{Winter:2008eg}.
For instance, even if there is no CP violation in $\deltacp$, one may discover CP violation from $\epsilon^m_{e\tau} \gtrsim 0.01$ in the neutrino propagation for up to 50\% to 80\% of all (true) values of the corresponding CP violating phase~\cite{Winter:2008eg}.  Whenever one discusses such large NSI, one should of course consider if such large NSI can be produced by an underlying gauge invariant model. In fact, one can show that a model producing such large $\epsilon^m_{e\tau}$ requires at least dimension eight effective operators with (at least) two mediator fields, satisfying a number of cancellation conditions~\cite{Gavela:2008ra}. This implies that the new physics scale cannot be much higher than about 300~GeV to discover such a new CP violation, and, because of the very specific model requirements, an arbitrary LHC-observable model will most likely not produce large enough effects.
Nevertheless, because it is impossible to predict the unexpected, one should follow an inclusive strategy, such as by using all available channels.

Another potentially unexpected case is the measurement of one of the relevant parameters by a different source of information, such as from a supernova explosion, from atmospheric neutrinos, LHC, or $0\nu\beta\beta$ decay.
As one example, flavor ratios in neutrino telescopes may allow for mixing parameter measurements if the accumulated statistics is good enough and the astrophysical sources are well enough known~\cite{Serpico:2005sz}. For example, the first measurement of $\deltacp$ could actually come from a reactor experiment plus astrophysical source in that case~\cite{Winter:2006ce}. Since source uncertainties affect the measurement and the accumulated statistics will probably be rather poor~\cite{Lipari:2007su}, one may choose pion decay sources only, for which the parameter dependence on $\deltacp$ is small in the standard oscillation scenario. In some neutrino decay scenarios, however, one can do the astrophysical plus reactor measurement even with this type of neutrino source or little knowledge on the source~\cite{Maltoni:2008jr}. This is only one out of many examples in which some external knowledge could
impact the long baseline strategy.

{\bf Summary and conclusions.}
In summary, choosing and optimizing a long baseline experiment faces a number of challenges. From the physics point of view, correlations and degeneracies demand resolution strategies, theory requires a number of performance indicators to be measured, and the potentially unexpected indicates to use strategies as inclusive as possible, such as to combine different channels. Apart from physics, other factors impact the experiment strategy, such as political support, regional interests, results from LHC, or possible measurements of the same quantities by different experiment classes. Because of these partially open boundary conditions, we should stay open for new possibilities, and develop the existing ones further.

\vspace*{-0.4cm}

{\footnotesize

} 

\end{document}